\documentclass{ws-procs9x6}

\newcommand{\Bdecay}{$ B \to X_c \,  \ell \,  \bar{\nu} $ }
\newcommand{\LQCD}{{\Lambda_{\rm QCD}}}
\newcommand{\orderalpha}{$ \mathcal{O}\!\left(\alpha_s \right) \,  $}
\newcommand{\orderalphalam}{$ \mathcal{O}\!\left( \alpha_s \, \LQCD / m_b
\right)  $ }

\begin{document}


\title{ Improving Extractions of $|V_{ cb}|$ and the $ b $ quark mass from Semileptonic Inclusive B decay }

\author{Michael Trott}
\address{Department of Physics, University of Toronto, \\ 60 St. George St.
	Toronto, Ontario, Canada, M5S 1A7 
mrtrott@physics.utoronto.ca}

\maketitle

\abstracts{
Recent advances in improving extractions of $|V_{cb}|$ and $m_b$ from spectra of 
semileptonic inclusive B decay are reported. Results of a general moment analysis 
of the lepton energy spectrum and the hadronic invariant mass spectrum  are summarized.
The  calculation of the general  \orderalpha structure functions for semileptonic B decay is reported,
which has allowed the calculation of the \orderalphalam terms for the hadronic invariant mass moments to be carried out. Recent theoretical advances and improvements in experimental data has allowed  extractions of the CKM matrix element  $|V_{cb}|$ to improve to the $2\% $ level. }


\section{Introduction}
Inclusive semileptonic \Bdecay decay offers the opportunity to measure the
CKM parameter $|V_{cb}|$ and the bottom quark mass from experiment.
\cite{ope,opecorr1,opecorr2,leptonmomentValoshin,leptonmoment4,leptonmoment2,hadronmoment2,hadronmoment3,hadron3rdorder}.
Experimental studies of moments of the differential decay spectra 
of \Bdecay, combined with a measurement of the total inclusive decay rate, 
allow the extraction of these parameters to occur with high precision. 

The idea is that sufficiently inclusive observables of this decay can be calculated without model dependence from QCD using an operator product expansion (OPE).
The OPE demonstrates that in the $m_b \gg \LQCD $ limit inclusive $ { \rm  B } $ decay rates are 
equal to inclusive $ {\rm b} $ quark decay rates.  Corrections to this rate are suppressed by powers of 
the ratio $\LQCD/ { m_b } $  and $\alpha_s $. The terms in the $\LQCD/ { m_b } $
expansion are parameterized using heavy quark effective theory (HQET) with
two parameters introduced for corrections at $ \mathcal{O}\!\left( \, \LQCD^2 / m_b^2
\right) $, typically labeled  $\lambda_{1,2}$, while six new parameters enter the expansion at 
$ \mathcal{O}\!\left( \, \LQCD^3 / m_b^3 \right)  $,  typically labeled $\rho_{1,2} $ and 
$\tau_{1,2,3,4} $. To extract  $|V_{cb}|$ from inclusive decay spectra with 
high precision one needs to accurately know the  $ {\rm b} $ quark mass and the 
largest terms in the OPE that the \Bdecay semileptonic decay rate depends upon.

Experimental results have been reported by the BABAR, BELLE, CDF, CLEO and DELPHI
collaborations measuring various $ { \rm B }$ meson decay spectra and moments\cite{CLEO2002,CLEO2004,CLEO20042,BABAR01,BABAR02,DELPHI02604,BELLE04}.
A number of authors have performed global fits to this data\cite{bllm,Battagliafit,CLEOfit,BABARfit},
the most recent of which\cite{bllmt} finds $|V_{cb}| = \left(41.3 \pm 0.6 \pm 0.1 \tau_B \right) \times 10^{-3}$ and 
$m_{b}^{1S} = 4.68 \pm 0.04 \, \rm{ GeV}$. 
Determining  $m_b$ and $|V_{cb}|$ to this level of precision, and prospects of further improvement, 
require that  the a high degree of confidence be obtained that the OPE fits the data and that the extraction of the unknown matrix elements of the leading order operators in the OPE be as accurate as possible. 

Recent work aimed at these goals has followed the approach of 
calculating general observables in order to maximize the amount 
of information obtained from experimental improvements in measured \Bdecay spectra\cite{BauerTrott2002,Trott2004}.  The calculation of the
\orderalpha structure functions for \Bdecay has also been completed\cite{Trott2004}
recently to reduce the theoretical uncertainty in hadronic invariant mass moments in 
determining these parameters.  

The purpose of this paper is to review recent work in this area; focusing in 
particular on basic features of the general moments analysis and
the calculation of the \orderalpha structure functions.

\section{General Moment Approach}\label{GeneralMoment}
A \Bdecay observable calculated by an OPE is a double expansion
in terms of the strong coupling $ \alpha_s (m_b) $  and 
the ratio $ \LQCD / m_b $. In obtaining predictions for observables 
the triple differential decay spectrum is decomposed in terms of the 
invariant mass of the $W$ boson 
$\hat{y} = q^2 / {m_b}^2 $, where  $q^{\mu}$ is the momentum of 
the lepton pair, the $c$ quark jet invariant mass 
$\hat{z} = {\left( m_b v - q \right)}^2 / {m_b}^2$,
and the charged lepton energy  $\hat{E}_{\ell} = E_{\ell}/m_b $. 

In extracting a b quark mass parameter,  the pole mass is related 
to the short distance 1S mass\cite{renorupsilon,upmassmanohar} 
to avoid the pole mass renormalon\cite{renormalon,luke_renormalon}
ambiguity that leads to unnecessarily badly behaved perturbation series.
We choose to express the $ {\rm b} $ quark pole mass perturbatively in terms 
of the 1S mass  and use the fact that 
$ \frac{m_{\Upsilon}}{2}-m_b^{\rm 1S} \sim \Lambda_{\rm QCD}$ 
to expand in the parameter $ \Lambda_{1S}  \equiv \frac{m_{\Upsilon}}{2} - m_b^{\rm 1S}\, $.

Following this approach we obtain predictions for the lepton energy 
spectrum and the hadronic invariant mass spectrum of \Bdecay.
Observables are usually calculated  with experimentally required cuts 
to reduce backgrounds,  such as a cut on the charged lepton energy,  
so that in general a prediction for an nth moment of an observable ($O^n$) is an expansion
of the following form

\begin{eqnarray}
& \, & \frac{1}{{\Gamma_0}}  \int_{\hat{E}_{\ell}^{min}} 
 O^n \left[ \hat{y} \, , \hat{z}  \, , \hat{E}_{\ell} \right] 
 \frac{d \Gamma}{d \hat{y} \, d \hat{z}  \,d \hat{E}_{\ell}}
 d \hat{y} \, d \hat{z}  \,d \hat{E}_{\ell}   
=  f_0 \left[ n , {\hat{E}_{\ell}^{min}} \right]   \\
& + & f_1 \left[ n , {\hat{E}_{\ell}^{min}} \right]  \frac {\Lambda_{1S}}{m_{\Upsilon}} 
 +  \sum _{i = 1}^2 f_{i+ 1} \left[ n ,  {\hat{E}_{\ell}^{min}}  \right]  \frac{ \lambda_i}{m_{\Upsilon}^2}  
+  \mathcal{O}\!\left( \alpha_s,  \frac{\LQCD^3 }{ m_{\Upsilon}^3} \right),  \nonumber
\end{eqnarray}
where
\begin{eqnarray}
\Gamma_0 = \frac{G_F^2 {|V_{cb}|}^2 \left( m_{\Upsilon} \right)^5}{192
\pi^3}.
\end{eqnarray}

The motivation of the general moment approach is to exploit
the degree of experimental and theoretical control that exists over
 $n$ and $ {\hat{E}_{\ell}^{min}} $ by examing directly the 
dependence of the various coefficient  functions $f_i$ on these terms. 
Using this control, one can uncover moments that are well suited to 
measure the HQET matrix elements and test our understanding of B decay.

\section{Lepton Energy Spectrum}\label{LeptonMoment}

\begin{figure}
\centerline{\epsfxsize=4.5in\epsfbox{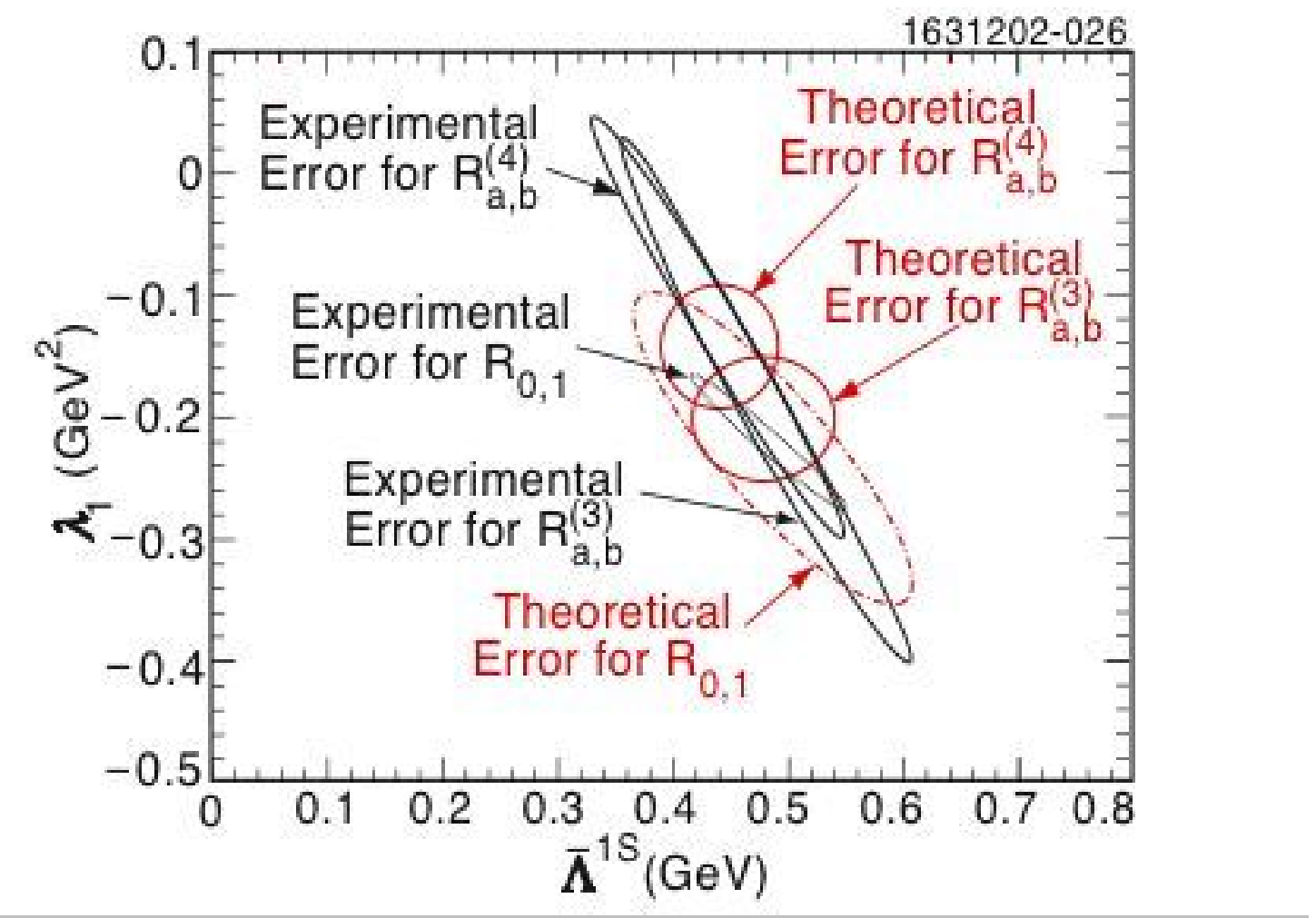}}  
\caption{Comparison of various pairs of moments in the lepton energy spectrum in extracting 
$\Lambda_{1S} $ and $ \lambda_1$. Although the theoretical error was reduced the
experimental error increased. From the CLEO collaboration, Mahmood { \it et all.}\label{momentsfig} }
\end{figure}

In the lepton energy spectrum, the ratio of lepton energy moments
was considered using this approach\cite{BauerTrott2002}
\begin{eqnarray}
R [n,E_{\ell 1},m,El_{\ell 2}] = \frac{\int_{E_{\ell 1}}^{m_B/2} E_\ell^n \frac{d\Gamma}{d E_\ell}}{\int_{E_{\ell 2}}^{m_B/2} E_{\ell }^m 
\frac{d\Gamma}{d E_\ell}}.     
\end{eqnarray}
The parameter space of $ n, m, E_{\ell 1}$ and $El_{\ell 2}$ was then examined. Moments were found with suppressed third order contributions in the nonperturbative expansion that improve the theoretical
error in the extractions of $m_b$, thru $\Lambda_{1S} $, and $ \lambda_1$. Sets of moments
of this type were found and have been experimentally examined as shown in Fig.~\ref{momentsfig}.  

Another possible source of theoretical error in extractions of $|V_{cb}|$ and $m_b$ is 
a quark-hadron duality violation effect that is not incorporated in the OPE analysis.
Some authors advocate that duality violating effects could be as large as order 
$ {\LQCD}/{m_{b}}$\cite{Isgur};  while other  authors argue that duality violations are small. 
As duality violation is difficult to reliably quantify theoretically, it is important to 
quantify duality violation as directly as possible from experiment.  

In global fits, duality violations effects could appear as a poor $ \chi^2 $, 
and no evidence has been found in recent fits for large duality violations\cite{bllmt}.  
These indirect tests of duality violation are limited by our  
knowledge of $m_b$ and $\lambda_1$ and recent fits have improved our knowledge of 
to $ \lambda_1 =  - 0.24 \pm 0.06 \, { \rm GeV^2 } $ and 
$m_{b}^{1S} = 4.68 \pm 0.04 \, \rm{ GeV}$.  

Another way to quantify the possible size of duality violating terms is to use a 
duality testing moment uncovered by examining the general lepton energy moment. 
This complementary technique is limited to a lesser degree by the error on $ \lambda_1 $ 
and  $m_b$: as a duality testing moment is defined to be a moment with the leading order 
unknown terms in the OPE suppressed.  Moments of this type can be predicted to percent level 
accuracy and the following table illustrates the impressive level of agreement that these moments 
exhibit with experimental data.

\begin{center}
TABLE 1: Lepton Energy Duality Moments \\   
\begin{tabular}{|c|c|c|c|c|}
\hline
\, & $D_3$ &  $D_4$  \\
\hline 
Theoretical Prediction & 0.5200 $\pm$ 0.0014 & 0.6053 $\pm$ 0.0018 \\
Experimental Results & 0.5193 $\pm$ 0.0008 & 0.6036 $\pm$ 0.0006 \\
\hline 
\end{tabular}
\end{center}

\section{Hadronic Invariant Mass Spectrum}\label{HadronMoment}

Calculating a general hadronic invariant mass moment presents unique challenges\cite{Trott2004}.
Proceeding as in the lepton energy case one can attempt to calculate 
\begin{eqnarray} \label{momentdefinition}
S [n,E_{\ell_1},m,E_{\ell_2}] = \frac{\int_{E_{\ell 1}} s_H^n\frac{d \Gamma}{d \hat{y} \, d \hat{z}  \,d \hat{E}_{\ell}}
 d \hat{y} \, d \hat{z}  \,d \hat{E}_{\ell} }{\int_{E_{\ell 2}} s_H^m 
\frac{d \Gamma}{d \hat{y} \, d \hat{z}  \,d \hat{E}_{\ell}}
 d \hat{y} \, d \hat{z}  \,d \hat{E}_{\ell} },
\end{eqnarray}
\noindent{where}
\begin{eqnarray}
s_H  = {m_B}^2 -  \, m_B \, m_b \, \left( 1 - \hat{z} +
\hat{y} \right) + m_b^2 \, \hat{y} .
\end{eqnarray}

However in order to perform the integrations required for the analysis
one has to expand the general moment of $s_H $ in the following manner,
\begin{eqnarray}
s_H^n =  \sum_{k = 0}^{\infty} \sum_{l = 0}^k \, \frac{\Gamma(n+ 1)}{ \Gamma(n+
1 - k)  \, \Gamma(k)} \, C^k_l \, \hat{y}^l \, \hat{z}^{n-k} \, m_b^{2 n} .
\end{eqnarray}
The coefficient functions $ C^k_l $ are $ O \left( \Lambda_{\rm QCD}^k /m^{k}_b
\right)$. Expanding up to $ O \left( \Lambda_{\rm QCD}^3 / m_b^3 \right)$
in the nonperturbative expansion one finds,

\begin{eqnarray}
s_H^n = \,\hat{z}^n \, m_b^{(2 \, n)} \,\Big[ C_0^0 + \frac{n}{\hat{z}} \left(
C_0^1 + \hat{y} \, C_1^1 \right)
+ \frac{n \, \left( n-1 \right)}{1! \,\hat{z}^2} \left(C_0^2 + \hat{y} \, C_1^2
+ \hat{y}^2 \, C_2^2 \right)  \nonumber \\
 + \frac{n \, \left( n-1 \right) \,  \left( n-2 \right)}{ 2! \,\hat{z}^3}
\left(C_0^3 + \hat{y} \, C_1^3 + \hat{y}^2 \, C_2^3 + \hat{y}^3 \, C_3^3
\right) \Big],
\end{eqnarray}
where the $C_l^k$ are functions of $n$  and the nonperturbative matrix elements.
For integer moments this expression has no $1/z$
dependence, however, for non-integer moments  one obtains
contributions of order $ z^{-k} $ where $k \ge n$ is the ceiling of the fractional moment
power $n$. As the lower limit of $z$ is $\rho = m_c^2  / {m_b^2}$  
this corresponds to a $ m_b \,  \LQCD / m_c^2 $  expansion entering into the calculations of fractional
moments. Formally this expansion is well behaved in the SV limit\cite{bllmt}  where 
$ m_b \sim m_c \gg m_b - m_c \gg \LQCD $. The precise manner to reliably estimate this
uncertainty is currently under study. 

In the hadronic invariant mass spectrum duality testing moments 
can also be found and the predictions of these moments should be compared to
experimental measurements. The hadronic invariant mass spectrum
also offers the opportunity to measure the $ { \rm b } $ quark mass with minimal 
error due to $ \lambda_1 $: as moments have been found that have a 
strong dependence on the $ {\rm b} $ quark mass, while having a suppressed 
dependence on $ \lambda_1 $. 

\section{\orderalpha Hadronic Structure Functions  }\label{OrderAlpha}

In determining  $|V_{cb} |$ and $m_b$,  the perturbative corrections 
to the hadronic invariant mass spectrum and the lepton energy spectrum 
are currently calculated  to $ \mathcal{O}\!\left( \alpha_s^2  \beta_0 \right) $.  
The $ \mathcal{O}\!\left( \alpha_s \right) $ spectra have been known
for some time\cite{JezabekKhun} and the  $ \mathcal{O}\!\left( \alpha_s^2  \beta_0 \right) $ corrections
are obtained from the $ \mathcal{O}\!\left( \alpha_s \right) $ result via a
dispersion integral.  As observables calculated in the OPE are a double expansion in the 
parameters $ \alpha_s $ and  ratio $\LQCD/ { m_b } $, the largest 
cross term of these expansions is the $ \mathcal{O}\! \left( \alpha_s  \LQCD / m_b \right) $
terms. Calculations of the $ \mathcal{O}\! \left( \alpha_s  \LQCD / m_b \right) $
terms in the lepton energy spectrum were obtained from the known 
$ \mathcal{O}\!\left( \alpha_s \right) $ spectra but the lepton energy cut dependence of the
$ \mathcal{O}\! \left( \alpha_s  \LQCD / m_b \right) $ terms was unknown for hadronic invariant mass observables until recently. 

\begin{figure}[ht]
\centerline{\epsfxsize=3.2in\epsfbox{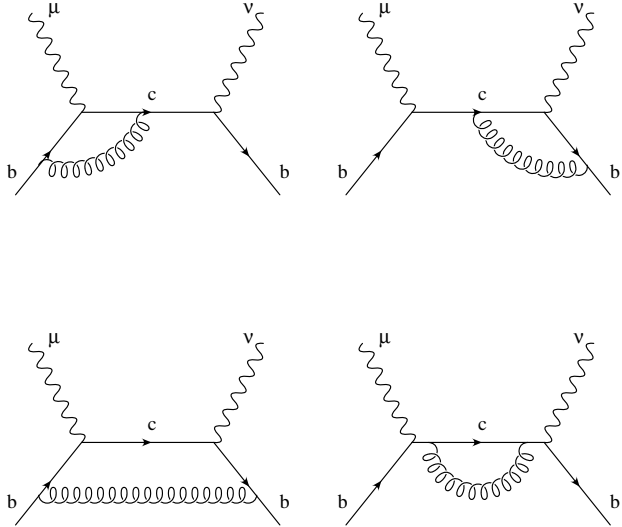}}   
\caption{The one loop forward scattering diagrams. The hadron tensor is derived
by calculating the imaginary part of the diagrams.\label{fig1}}
\end{figure}

This lack of  knowledge of the lepton energy cut dependence of the  $ \mathcal{O}\! \left( \alpha_s  \LQCD / m_b \right) $ terms was the largest theoretical uncertainty in hadronic invariant mass observables.  In order to determine this dependence, the $ \mathcal{O}\!\left( \alpha_s \right) $ corrections to the structure functions for a massive final state had to be determined;
which required a systematic calculation of cuts across all intermediate state contributions, while keeping the final state mass scale, to the diagrams shown in Fig.~\ref{fig1}.  This challenging calculation has recently been completed\cite{Trott2004} and the $ \mathcal{O}\!\left( \alpha_s \right) $ corrections to the structure functions for a massive final state are now known.
As a check of this calculation,  the massless limit of all regular terms of the $ \mathcal{O}\!\left( \alpha_s \right) $ structure functions  was taken and found to be in agreement with the $ \mathcal{O}\!\left( \alpha_s \right) $ contributions to the structure functions for a massless final state, which have been known for some time\cite{NeubertdeFazio}.  

It deserves to be emphasized that the 
$ \mathcal{O}\!\left( \alpha_s \right) $ corrections to all inclusive \Bdecay observables, 
with arbitrary cuts on kinematic variables, can now be determined in a systematic fashion.  
In particular, these general results were used to determine  
the lepton energy cut dependence of the  $ \mathcal{O}\! \left( \alpha_s  \LQCD / m_b \right) $ terms for the hadronic invariant mass spectrum, improving the agreement of these 
theoretical expressions with the data and removing the largest theoretical uncertainty
in these observables.  


\section{Conclusions}

The determination of $|V_{cb} | $ with a theoretical uncertainty below the 2\% level is a significant
theoretical achievement. The work reported on in this note represents a part 
of the efforts of a large number of theorists over the past decade and a half in
developing and applying the OPE techniques required in this extraction. Improvements in extracting 
other CKM matrix elements, in particular $ |V_{ub} |$, to this level of precision, 
will allow the consistency of the CKM description of CP violation to be precisely tested in the 
near future.
 

\section*{Acknowledgments}
 It is a pleasure to acknowledge Christian Bauer for collaboration on aspects of the 
 work reported on here. I am also grateful to Michael Luke, Craig Burrell and particularly 
Alex Williamson for numerous helpful discussions over the course of this work.

This work was supported in part by the Walter B. Sumner foundation.


\end{document}